# Design of Multifunctional Soft Doming Actuator for Soft Machines

*Yichao Tang and Jie Yin[*]*

Y. Tang, Prof. J. Yin.
Applied Mechanics of Materials Laboratory
Department of Mechanical Engineering
Temple University
1947 North 12th Street, Philadelphia, PA 19122, USA
E-mail: jieyin@temple.edu

**Abstract**: Bilayer bending based soft actuators are widely utilized in soft robotics for locomotion and object gripping. However, studies on soft actuators based on bilayer doming remain largely unexplored despite the often-observed dome-like shapes in undersea animals such as jellyfish and octopus's suction cup. Here, based on the simplified model of bending-induced doming of circular bilayer plates with mismatched deformation, we explore the design of soft doming actuator upon pneumatic actuation and its implications in design of multifunctional soft machines. The bilayer actuator is composed of patterned embedded pneumatic channel on top for radial expansion and a solid elastomeric layer on bottom for strain-limiting. We show that both the cavity volume and bending angle at the rim of the actuated dome can be controlled by tuning the height gradient of the pneumatic channel along the radial direction. We demonstrate its potential multifunctional applications in swimming, adhesion, and gripping, including high efficient jellyfish-inspired underwater soft robots with locomotion speed of 84 cm/min and rotation-based soft grippers with low energy cost by harnessing the large rim bending angle, as well as octopus-inspired soft adhesion actuators with strong and switchable adhesion force of over 10 N by utilizing the large cavity volume.

## 1. Introduction

The design of soft actuators, a key part in soft robotics for deformation actuation, has recently attracted tremendous research interest due to their broad applications in programmable locomotion[1], artificial muscles[2], and soft grippers[1a, 3]. A variety of soft actuators have been designed and fabricated to achieve different deformation modes including contraction/expansion[4], twisting[5], rotation[6], and bending[7]. Among them, bilayer bending based soft actuators are widely used for bending deformation actuation. The bilayer soft actuator is often constructed by bonding a strain-limiting layer to stimuli-responsive expanding structures. When in response to external stimuli such as pneumatic/hydraulic pressure[1a], light[5a], humidity[8], electrical[4a] and magnetic field[9], mismatched deformation generated in the bilayer structure will lead to the bending of the actuator. These bending actuators have been widely utilized to build functional soft robotics with capabilities of object manipulation[3, 10], locomotion[1b, 11], and assisting rehabilitation[12].

Despite the advance, the deformation in most of the soft bilayer bending actuators is limited to the bending in one direction while leaving the other orthogonal direction unbent, thus the deformed configuration often takes an open shape with zero Gaussian curvature after actuation, which may limit their applications to certain situations when enclosed actuated configurations with non-zero Gaussian curvature are needed. When bending in both orthogonal directions is allowed, it is known that a circular bilayer plate can bend or buckle into a dome-like shape with positive Gaussian curvature upon mismatched deformation between the two layers[13]. Similar dome-like shapes are often found in undersea animals. For example, a jellyfish bends its dome-shaped soft body for locomotion under the sea, where bending deformation expulses the water inside the dome to propel itself forward through contracting and relaxing the muscles around the dome[14]. Another example is the dome-like suction cups on the arms of octopuses for gripping and moving around through muscle contraction to generate negative pressure inside the chamber[15]. However, the potential applications of dome-like bending actuators in design of soft machines remains largely unexplored.

Here, we propose a new soft doming actuator consisting of patterned pneumatic channels on top and strain-limiting layer underneath. Upon pneumatic actuation, the bilayer circular planar structure can reversibly bend into a 3D dome-like shape. Based on the simplified bilayer bending model of circular plates[13], we explore the mechanics-guided design of controllable deformation in a soft doming actuator for its potential multifunctional applications in soft robotics.

**2. Results and Discussion**

**2.1. Design of a bilayer doming actuator for multifunctionality**

As schematically illustrated in Figure 1a, the proposed soft doming actuator is composed of a circular bilayer system with embedded patterned pneumatic spiral channels on the top (blue color) and solid elastomeric layer underneath (yellow color) for strain-limiting purpose. The actuator is made of soft silicone rubber, Ecoflex 00-50 (Smooth-on Inc) (top of Figure 1b and experimental section). Upon inflating air into the spiral pneumatic channel, the top-layer expansion along the radial direction renders a mismatched deformation between the top and bottom layer, thus forms a 3D dome-like shape (bottom of Figure 1b). After depressurization, the dome shape returns to its planar bilayer structure.

Guided by the simplified bending model of linear elastic circular-shaped bilayer structures with nonuniform axisymmetric mismatched strain (discussed in Section 2.2), we demonstrate the controllable deformation in the soft doming actuator to achieve either large cavity volume or large bending angle at its rim, as well as its potential applications for multifunctional soft machines, including bio-inspired design of a jellyfish-like soft underwater robot with high locomotion velocity (Figure 1c, discussed in Section 2.2), a soft adhesion actuator with strong and switchable adhesion force (Figure 1d, discussed in Section 2.3), as well as a soft gripper with low energy cost (Figure 1e, discussed in Section 2.4).

## 2.2. Simplified theoretical modeling

To shed some light on the design of bilayer-doming based soft robotics, we employ a simplified bilayer model with nonuniform axisymmetric mismatched expansion between two layers to understand the deformation of the bilayer doming structure, particularly the dome height and volume for design of adhesion actuator, and the bending angle at the rim for design of underwater swimmers with high thrust force and rotation-based soft grippers.

For a bilayer system composed of a circular thin film (thickness of $h_f$) on a substrate (thickness of $h_s$) with radius of $R$ as shown in Figure 2a ($h_f \ll h_s$), when it is subjected to a nonuniform but axisymmetric misfit strain $\varepsilon_m(r)$ along the radial direction $r$, the height $u_z$ of the deformed dome structure along the normal direction $z$-axis can be obtained as[13b]

$$\frac{du_z}{dr} = 6\frac{E_f h_f}{1-v_f^2}\frac{1-v_s^2}{E_s h_s^2}\left[(1+v_s)\frac{1}{r}\int_0^r \eta\varepsilon_m(\eta)d\eta + (1+v_f)\frac{1-v_s}{1+v_s}\frac{r}{R^2}\int_0^R \eta\varepsilon_m(\eta)d\eta\right] \quad (1)$$

where $E$ and $v$ are the Young's modulus and Poisson's ratio. The subscripts "$f$" and "$s$" represent the film and substrate, respectively. $r$ is defined as the radial distance from the center. The slope or bending angle of the dome at the rim, which we call the "doming angle" $\varphi$ in Figure 2b, can be obtained as:

$$\varphi = \frac{du_z}{dr}(R) = 6\frac{E_f h_f}{1-v_f^2}\frac{1-v_s^2}{E_s h_s^2}\left[(1+v_s)\frac{1}{R}\int_0^R \eta\varepsilon_m(\eta)d\eta + (1+v_f)\frac{1-v_s}{1+v_s}\frac{1}{R}\int_0^R \eta\varepsilon_m(\eta)d\eta\right] \quad (2)$$

By integrating Equation (1) with respect to $r$, we can get the dome height $u_z$:

$$u_z = 6\frac{E_f h_f}{1-v_f^2}\frac{1-v_s^2}{E_s h_s^2}\left[(1+v_s)\int_0^r \frac{1}{r}\int_0^R \eta\varepsilon_m(\eta)d\eta dr + (1+v_f)\frac{1-v_s}{1+v_s}\frac{r^2}{2R^2}\int_0^R \eta\varepsilon_m(\eta)d\eta\right] + C \quad (3)$$

where $C$ is a constant to be determined by satisfying the assumed boundary condition of $u_z(R) = 0$.

It should be noted that when the model is applied to understand the deformation in the doming actuator, it is oversimplified by homogenizing the expanding layer without considering its detailed patterned pneumatic channels. The misfit strain $\varepsilon_m$ between the two layers is governed by the pneumatic channeled structure, which can be manipulated by controlling the geometry of the spiral channel along the radial direction. After homogenization, we assume that $E_f \approx E_s$ and $v_f \approx v_s$. Equation (2) and (3) show that for a bilayer plate system with given geometry, i.e. normalized layer thickness $h_f/R$ and $h_s/R$, the dome height and rim slope are mainly determined by the axisymmetric misfit strain $\varepsilon_m$ controlled by the channel geometry. In the following sections, based on this simplified model, we will demonstrate that by manipulating the channel height gradient along the radial direction, the nonuniform misfit strain $\varepsilon_m(r)$ can be tuned to allow more expansion either near the center or the rim to achieve large cavity volume or large dome bending angle at the rim in the actuated dome shape. The different features of the deformed dome shapes will be utilized in soft doming actuators to guide the design of multifunctional soft robotics as discussed below.

**2.3. Swimming actuator**

The reversible switch from flat to dome-like shape in the doming actuator upon pressurization is similar to the deformation of jellyfish body, which inspires us for exploring its potential application in design of jellyfish-like underwater soft robots. The proposed swimming robot is schematically shown in Figure 3a. It is composed of a pneumatic doming actuator made of soft silicone Ecoflex 00-50 and attached with a few stiff plastic film-based propellers to amplify the thrust force under water. Figure 3b illustrates the representative underwater locomotion modes. At rest state, the soft actuator remains flat and undeformed (left of Figure 3b). Upon inflation, it bends into a dome shape. The dome contract upon bending expulses water to push the swimmer forward. Meanwhile, the attached propellers flap backward correspondingly upon dome bending to generate vortex under water for enhancing the thrust force. Upon deflation, it returns to its flat state. During swimming, for the dome shape with given size, the doming angle at the rim plays a dominant role in determining the thrust force and thus the swimming speed, as evidenced by the locomotion of jellyfishes [16]. The high locomotion efficiency in jellyfish results from not only the abrupt muscle contraction[14], but from a large bending angle at the rim of its dome-shaped body that can generate larger vortices to propel it forward. Therefore, bioinspired by the performance of the jellyfish, we exploit the manipulation of $\varepsilon_m(r)$ in the doming actuators to achieve a relatively larger doming angle, as well as examine their underwater performance for designing potential high-efficient underwater soft robots.

According to Equation (2), for a bilayer actuator with given geometry, the doming angle $\varphi$ is mainly determined by the misfit strain $\varepsilon_m(r)$ along the radial direction, which can be tuned by varying the height of the channel along the radial direction, as shown in Figure 3c. In general, a deep or high aspect-ratio pneumatic channel yields a larger in-plane expansion than the shallow one. By tuning the height gradient of the pneumatic channel along the radial direction, i.e. the value of the tilting angle $\theta$ defined in Figure 3c, we can manipulate the nonuniform radial expansion of the pneumatic layer and thus the doming angle $\varphi$ at the rim. Here, a negative value of $\theta$, i.e. $\theta < 0$, denotes that the channel height decreases linearly from the edge to the center and presents a larger in-plane expansion close to the edge; a positive $\theta$, i.e. $\theta > 0$, indicates an increasing channel height from the edge to the center and presents a larger expansion close to the center; while $\theta = 0$ represents a constant channel height in the top layer.

To understand the relationship between $\theta$ and $\varepsilon_m(r)$, we use the digital image correlation (DIC) to track the expansion of the top layer (indicated by blue in Figure 3c) and thus quantify the value of $\varepsilon_m$ as a function of $r$ upon inflation for doming actuators with different values of $\theta$. In the test, all the samples have the identical geometry ($R = 38$ mm, $h_s = 9$ mm, $h_f$

= 1 mm, $h_c$ = 5 mm) and are inflated with the same air input volume of 4mL. Figure 3d shows the measured expansion rate as a function of radial distance $r/R$ for three representative values of $\theta$ (i.e. $\theta = 1.5°$, $\theta = 0°$ and $\theta = -1.5°$). It shows that for different small value of $\theta$, the radial expansion in both the center and the rim (i.e. $r/R=0$) of the dome is close to zero, and it exhibits a peak value between them. As $\theta$ decreases from a positive value to a negative one, the position of the peak radial expansion rate shifts from close-to-center to close-to-rim, i.e. more expansion at the rim than in the center, which is consistent with the expectation of the gradient channel height.

Equipped with the information of measured $\varepsilon_m(r)$, next, we evaluate the corresponding doming angle in terms of the theoretical model in Equation (2). After substituting the fitted experimental curves of $\varepsilon_m$ in Figure 3d into Equation (2) (equations of fitted curves can be found in Figure S1), we can get the theoretical value of the doming angle $\varphi = 5.02°$, 4.46° and 3.90° for $\theta = -1.5°$, $\theta = 0°$ and $\theta = 1.5°$, respectively. It indicates that a negative $\theta$ will contribute to a larger doming angle $\varphi$ at the edge when compared to its counterparts with positive $\theta$, which is consistent with the expectation that a negative $\theta$ leads to a larger expansion close to the edge and thus a larger bending angle at the edge. Then we compare the results from the theoretical model with the experiments and we find that the theoretical value of $\varphi$ is moderately lower than the corresponding measured value shown in Figure 3e due to the oversimplified model. However, the model well captures the trend of the measured value of $\varphi$ with $\theta$ for different air input volume, where the actuator with a negative value of $\theta$ exhibits a moderately higher $\varphi$ than their counterparts with positive and zero value of $\theta$ and such a disparity increases with the expansion rate (Figure 3e).

Next, based on the knowledge of controlling the doming angle, we utilize the soft doming actuator to design of jellyfish-like underwater soft robots with a relatively high swimming speed. As discussed before, we expect that a large bending angle at the edge will yield a large thrust force underwater. To validate it, we build three soft underwater swimmers with $\theta = -1.5°$, $\theta = 0°$, and $\theta = 1.5°$ and perform tests to measure their respective locomotion speed (Figure 3f). All the actuators have the same geometry ($R = 38$mm, $h_s = 9$mm, $h_f = 1$mm, $h_c = 5$mm) and are actuated at the same pressure (30kPa) and the same average frequency (0.333Hz). Figure 3g shows the image snapshots of the actuator taken at 2s, 7s, and 23s during swimming in the water tunnel (Movie S1). It shows that the actuator with $\theta = -1.5°$ can achieve the fastest average locomotion speed of 84 cm/min at the average actuation frequency of 0.333Hz (Figure S2) with the help of its relatively larger doming angle (Figure 3e). This is consistent with our expectation that a negative height gradient $\theta$ allows for more expansion around the edge to generate a larger bending angle at the edge and thus a larger thrust force.

It should be noted that despite previous studies of jellyfish-inspired soft robots based on different actuators such as ionic polymer metal composite (IPMC)[16a], shape memory alloy[16b], and dielectric elastomer[17], the proposed swimming robot based on the bilayer doming actuator is simpler in both structures and materials, and does not require complicated manufacturing process to achieve an even higher swimming speed than that of most reported active materials based jellyfish-inspired underwater robots[17-18], which demonstrates its potential advantage in designing high-efficient underwater robots.

### 2.4. Switchable adhesion actuator

In addition to the demonstration as a potential underwater soft robot, the similar dome-like shape in the bilayer doming actuator as the suction cup of octopuses inspires us for exploring its multifunctionality as a potential adhesion actuator[19].

Figure 4a schematically illustrates the working mechanism as an adhesion actuator. When attached to a foreign surface, upon pneumatic pressurization on the top layer, the planar circular bilayer structure will continuously "pop up" and deform into a dome-like shape upon radial expansion of the top layer, leaving a cavity with high vacuum between the popped-structure and the attached surface. The pressure difference between the cavity and the outer circumstance will force the actuator to firmly adhere to the target surface. Upon depressurization, the deformed dome-like shape will return to its original planar configuration for easy detachment.

Different from the design principle of achieving a larger doming angle in underwater soft robot discussed in Section 2.3, the goal for designing adhesion actuators is to achieve a high vacuum in the cavity for a large adhesion force by maximizing the volume of the cavity after deformation. The cavity volume is mainly determined by the dome height $u_z$. Thus, a positive value of channel height gradient $\theta$ is preferred as shown in Figure 4b. Compared to actuators with negative $\theta$, a positive $\theta$ allows a larger expansion close to the center than at the edge to achieve a larger dome height in the center. Furthermore, as discussed above, as $\theta$ ($\theta>0$) increases, the peak expansion will shift closer to the center, which generally will lead to a larger dome height and thus a larger adhesion force.

To examine the design principle, we fabricate three adhesion actuators with different values of $\theta$ (i.e. $\theta = 0°$, $\theta = 1.5°$, and $\theta = 3°$) while keeping the other geometrical sizes the same (i.e. $R = 28$mm, $h_s = 9$mm, $h_f = 3$mm, $h_c = 5$mm). Figure 4c shows the corresponding measured radial expansion rate from the center to the edge as a function of $\theta$ through DIC under the same 4mL air input. It shows that as $\theta$ increases, the peak expansion does shift closer to the center. Similarly, after substituting the fitted curves of the radial expansion rates in Figure 4c (the equations for fitted curves can be found in Figure S3) into Equation (3), the theoretically predicted profiles of the deformed dome shape can be obtained for different $\theta$, which is shown in Figure 4d. It shows that as $\theta$ increases from 0° to 3°, the dome height at the center ($r/R = 0$) increases slightly.

Next, we examine the adhesion strength of the three soft adhesion actuators with different $\theta$ by measuring the normal adhesion force on a smooth acrylic surface. The adhesion strength of the actuator is quantified by measuring the maximum normal adhesion force through the pulling force testing as illustrated in Figure 4e. The measured adhesion strength as a function of $\theta$ is shown in Figure 4f. It shows that the maximum adhesion force increases approximately linearly with $\theta$ and becomes almost doubled as $\theta$ increases from 0º to 3º.

However, such a largely increased adhesion strength with $\theta$ observed in experiments does not agree well with the theoretical model, where a small increase in the adhesion strength with $\theta$ is predicted due to the slightly increased cavity volume. The disparity results from the different deformation mechanisms for open and close bilayer dome structures. For the open bilayer dome actuator without attaching to a surface, i.e. the case of simplified model, the structural deformation is mainly determined by the mismatched expansion of the top layer induced bilayer bending, i.e. a "pop-up" deformation. However, when attaching to a surface, the suction force resulting from the pressure difference is absent in the open dome and not considered in the simplified model. The suction force intends to pull down the "pop-up" structure, thus generating a potential "bi-stable" dome structure, depending on the competition between the "pull-up" force arising from the expansion-induced bending and the "pull-down" force arising from the pressure difference in the cavity.

In experiment, we do observe the deformation bifurcation in the adhesion actuator (Figure S4). As the top layer starts to expand, the bilayer structure initially deforms into an axisymmetric dome-shape. However, as the pressure difference between the cavity and the ambient environment builds up upon further expansion, when beyond certain critical point, bifurcation may occur and break the axial symmetry of the dome structure to generate an asymmetric dome shape as shown in the left inset of Figure 4f and Figure S4. This distorted configuration may weaken the adhesion behavior of the soft adhesion actuator upon further pressurization. We note that despite the observed bifurcation in the bi-stable bilayer doming system, a relatively larger value of height gradient $\theta$ ($\theta > 1.5º$), i.e. more radial expansion in the center, can help to delay the bifurcation and hold the axial-symmetric dome-shape configuration even at a large mismatch strain without localized structural collapse, thus to enhance the large adhesion force even at a large actuation pressure of 40 kPa (Figure 4f and right inset). In a contrast, adhesion actuators with smaller angle (e.g. $\theta = 0.45°, 0.9°$ and $1.35°$) deform into distorted shapes (left inset of Figure 4f) and demonstrate smaller adhesion force at the same actuation condition.

**2.5. Gripping actuator**

The observed large bending angle at the edge of the bilayer dome structure enables the design of a potential gripping actuator by harnessing the controllable bending-induced rotation of attached gripper arms for object pick-up and drop-off. As discussed in Section 2.3, a larger expansion close to the edge than around the center in the top layer is preferred to achieve a large bending angle $\varphi$ at the edge. Thus, to further enhance $\varphi$ for design of gripping actuators, we propose a modified design of an annulus-shaped bilayer plate as schematically illustrated in Figure 5a, where the central part of the original solid bilayer plate-based actuator is cut out with a radius of $R_{in}$. Similarly, the top expansion layer is embedded with pneumatic spiral channels of the same height, while the layer underneath (yellow color) does not expand for strain-limiting purpose. The corresponding simplified homogenization bilayer model is shown in Figure 5b. Compared to its counterpart without cut-out, the annulus-shaped bilayer actuator has two potential benefits: one is to achieve a larger $\varphi$ by manipulating the size of the cut-out and allowing more expansion shifting to the outer annulus boundary; the other is to reduce the energy cost to realize the same bending angle $\varphi$ without the need to bend the original top cap region.

To reveal the geometrical effect on the bending angle $\varphi$ of the annulus bilayer structure, some useful insights can be obtained from the theoretical model on the deformation of annulus bilayer plates with mismatched expansion stain $\varepsilon_m$ between the bilayer. The height $u_z$ of the deformed dome structure along the normal direction $z$-axis can be obtained as

$$\frac{du_z}{dr} = 6\frac{E_f h_f}{1-v_f^2}\frac{1-v_s^2}{E_s h_s^2}\left[(1+v_s)\frac{1}{r}\int_{R_{in}}^{r}\eta\varepsilon_m(\eta)d\eta\right]+Ar+B \tag{4}$$

The value of constants $A$ and $B$ can be obtained through the boundary conditions (see the Supporting Information for details). Then the doming angle can be obtained as:

$$\varphi = \frac{du_z}{dr}(R) = 6\frac{E_f h_f}{1-v_f^2}\frac{1-v_s^2}{E_s h_s^2}\left[(1+v_s)\frac{1}{R}\int_{R_{in}}^{R}\eta\varepsilon_m(\eta)d\eta\right]+AR+B \tag{5}$$

By assuming an approximately constant mismatched strain $\varepsilon_m$ in the annulus bilayer actuator with the same channel height, the theoretical prediction of $\varphi$ for annulus bilayer structures with different size of circular cut-out (i.e. $R_{in}/R$) is plotted in Figure 6a. It shows that at the same mismatch strain (e.g. $\varepsilon_m = 0.2$), the doming angle $\varphi$ increases almost linearly with $R_{in}/R$, which means that the actuators with a larger cut-out radius $R_{in}$ will result in a larger bending angle at its outer boundary. For example, for actuators with $R_{in}/R = 0.5$, $\varphi$ could reach a large value of over 60°.

To validate the model, we build the modified bilayer doming actuators with the same size of the outer radius of $R = 38$ mm but with different cut-out sizes $R_{in}$. The other geometrical sizes are kept the same ($h_s = 9$ mm, $h_f = 1$mm, $h_c = 5$ mm). Upon the same actuation pressure of 30 kPa, Figure 6b shows that as $R_{in}/R$ increase from 0 (i.e. no cut-out) to 0.5, correspondingly, $\varphi$ increases monotonically from ~ 41° to ~ 51°, which is consistent with the model and the expectation that the introduction of cut-out to the solid circular bilayer structure can help to enlarge the bending angle at the same actuation pressure.

Based on the improved doming angle of the modified bilayer doming actuator, next, we apply it to design a simple proof-of-concept soft pneumatic gripper. As illustrated in Figure 6c, the gripper is composed of an annulus pneumatic bilayer soft actuator with three 3D printed plastic gripping assistors attached to its edge. As the bilayer doming actuator bends up upon pressurization, the attached gripper arms will rotate correspondingly toward the center and close its arms to pick up the object. The proof-of-concept experiment shows that the built gripper actuator ($R_{in}/R = 0.5$, $R = 38$ mm, $h_s = 9$ mm, $h_f = 1$ mm, $h_c = 5$ mm) can effectively grasp and release the object (e.g. a plastic cup) by simply pressurizing and depressurizing the pneumatic channel with a small pressure of 30 kPa as shown in Figure 6d and Movie S2.

## 3. Conclusion

In summary, we demonstrate that by controlling the mismatched expansion in a simple circular bilayer system, the generated dome-like structure can yield (i) large-volume cavity and (ii) large doming angle at the edge, which can be utilized to develop multifunctional soft robots with capabilities of swimming, adhesion and grasping. This study serves as a guideline for designing doming-based soft robots. We believe that the design principle of harnessing mismatched deformation for designing doming-based actuators could be applied to not only elastomeric materials such as the silicone rubber in this study or hydrogels actuated by hydraulic pressure, but other stimuli-responsive materials such as liquid crystals, shape memory polymer, and dielectric elastomers in response to light, heat, and electric field etc. The doming-based bending actuator could find broad potential applications in design of multifunctional soft machines such as underwater swimmer, climbing soft robots by harnessing the switchable adhesion[19c, 20], and jumping soft robots by harnessing the bistable characteristics of the dome structure[21] etc.

**Experimental Section**

*Actuator fabrication*: All pneumatic doming actuators were fabricated following the typical manufacturing technique for fluid-driven soft actuators reported by Ref. [1a]. Ecoflex 00-50 (Smooth-on Inc) was used for both pneumatic channeled layer and the strain limiting layer. The two layers were directly cast from molds printed by Ultimaker 2+ separately and were cured at 70°C for two hours. Then we glue the two layers together with Ecoflex 00-50 and cure them at 70°C for another 1 hour.

*Adhesion measurement*: The normal adhesion force of the doming actuator was measured using Instron 5944 with a 2kN load cell. The soft actuators were pressurized at 40kPa and the extension rate of the Instron was 1 mm min$^{-1}$.

*Digital Image Correlation (DIC) characterization*: Speckles were sprayed on the top surface of the soft actuator using an airbrush and India ink for DIC measurement. Images of the testing were taken at a rate of 1 fps (VicSnap, Correlated Solution) and DIC (Vic-2D, Correlated Solution) was used to track the deformation and obtain local strain contours.

**Acknowledgements**
J. Y. acknowledges the financial support from the start-up fund at College of Engineering at Temple University.

**Figures and Captions**

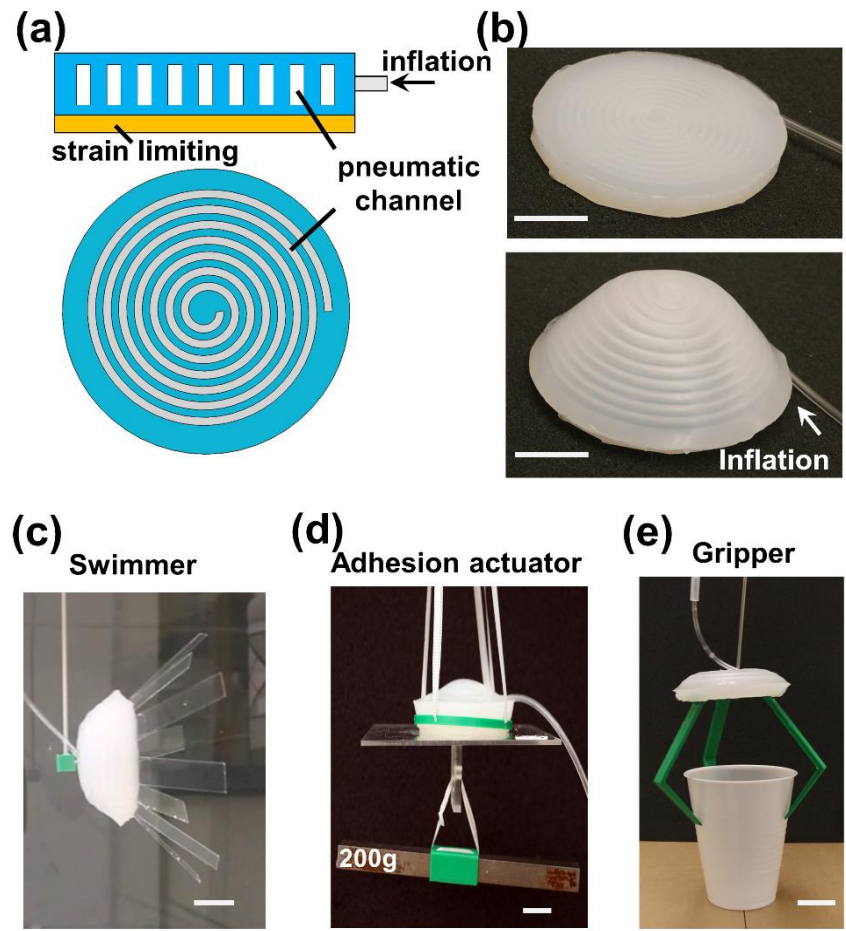

**Figure 1.** Design of multifunctional soft doming actuators. (a) Schematic of the circular bilayer doming actuator with embedded spiral-shape pneumatic channel on top and solid strain-limiting layer on bottom. (b) The as-fabricated doming actuator deforms into a dome-shape upon pressurization in the air channel. (c)-(e) Demonstration of soft doming actuators-based soft robots with capabilities of (c) swimming, (d) adhesion, and (e) grasping. The scale bar is 20mm.

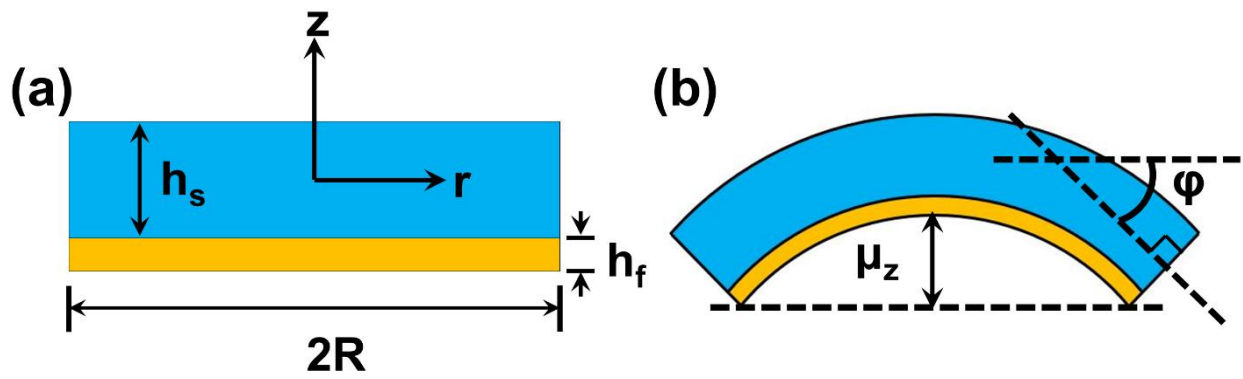

**Figure 2.** Schematic on the geometry of simplified homogenized circular bilayer systems before (a) and after actuation (b). The top layer denotes the expansion layer and the bottom one represents the strain limiting layer.

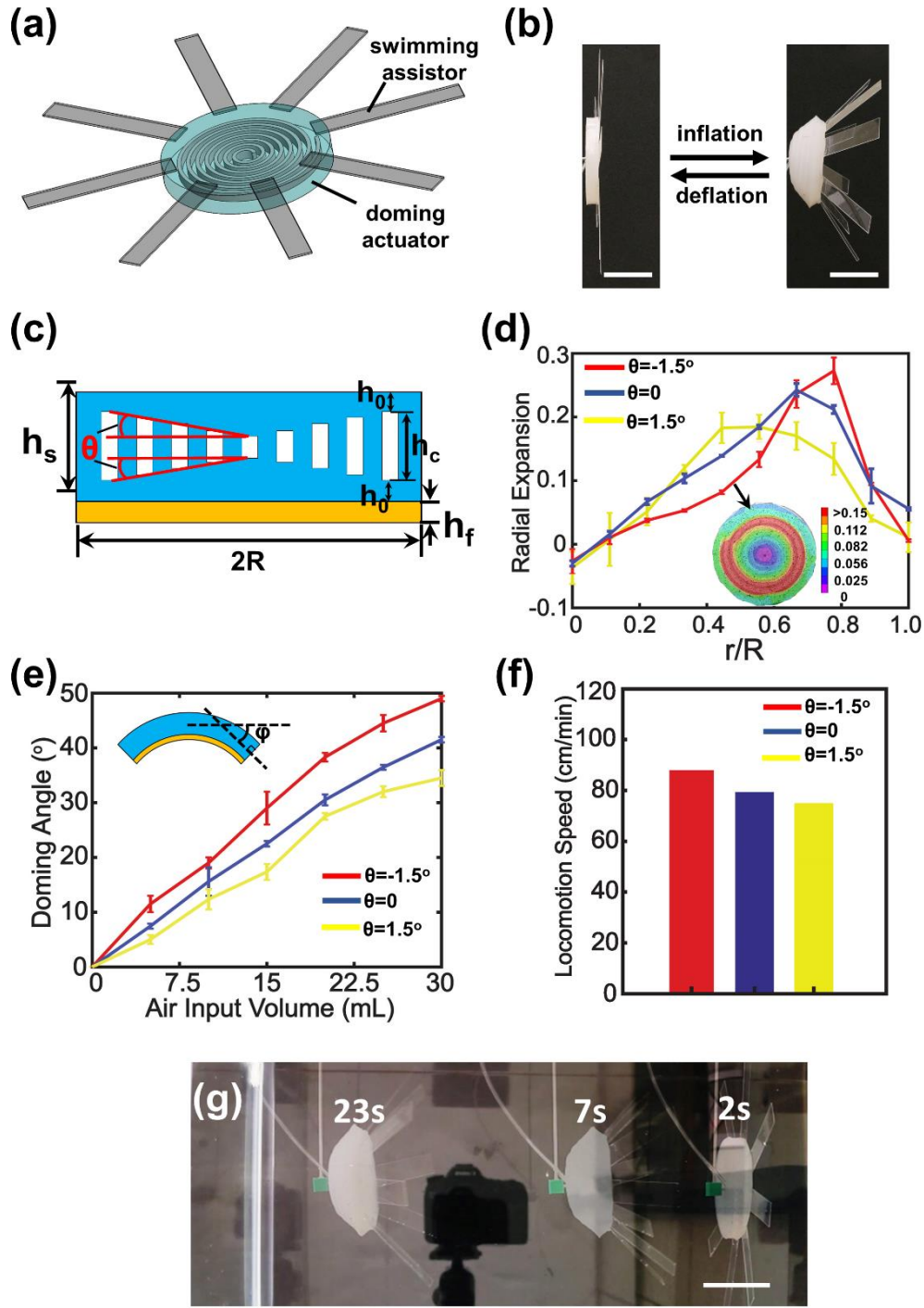

**Figure 3.** Design of jellyfish-inspired swimming soft robot based on the bilayer doming actuator. (a) Schematic design of swimming robot composed of a doming actuator (with spiral pneumatic channel) and polymeric swimming assistors (green) around the edge. (b) Locomotion modes of the proposed soft robot at rest state (left) and upon actuation (right). (c) Schematic illustration of bilayer doming model with spiral channel height gradient of $\theta$ along

the radial direction. (d) DIC measured radial expansion rate of the actuators with $\theta$=-1.5º, $\theta$=0º and $\theta$=1.5º upon 4mL inflation. The inset shows the radial strain contour in the actuator with $\theta$=-1.5º. (e) The measured doming angle at the edge as a function of air input volume for doming actuators with $\theta$=-1.5º, $\theta$=0º and $\theta$=1.5º. (f) Comparison in swimming velocity of proposed swimming robots with $\theta$=-1.5º, $\theta$=0º and $\theta$=1.5º. (g) Demonstration of the underwater locomotion of the proposed swimming robot at different actuation time. The scale bar is 50mm.

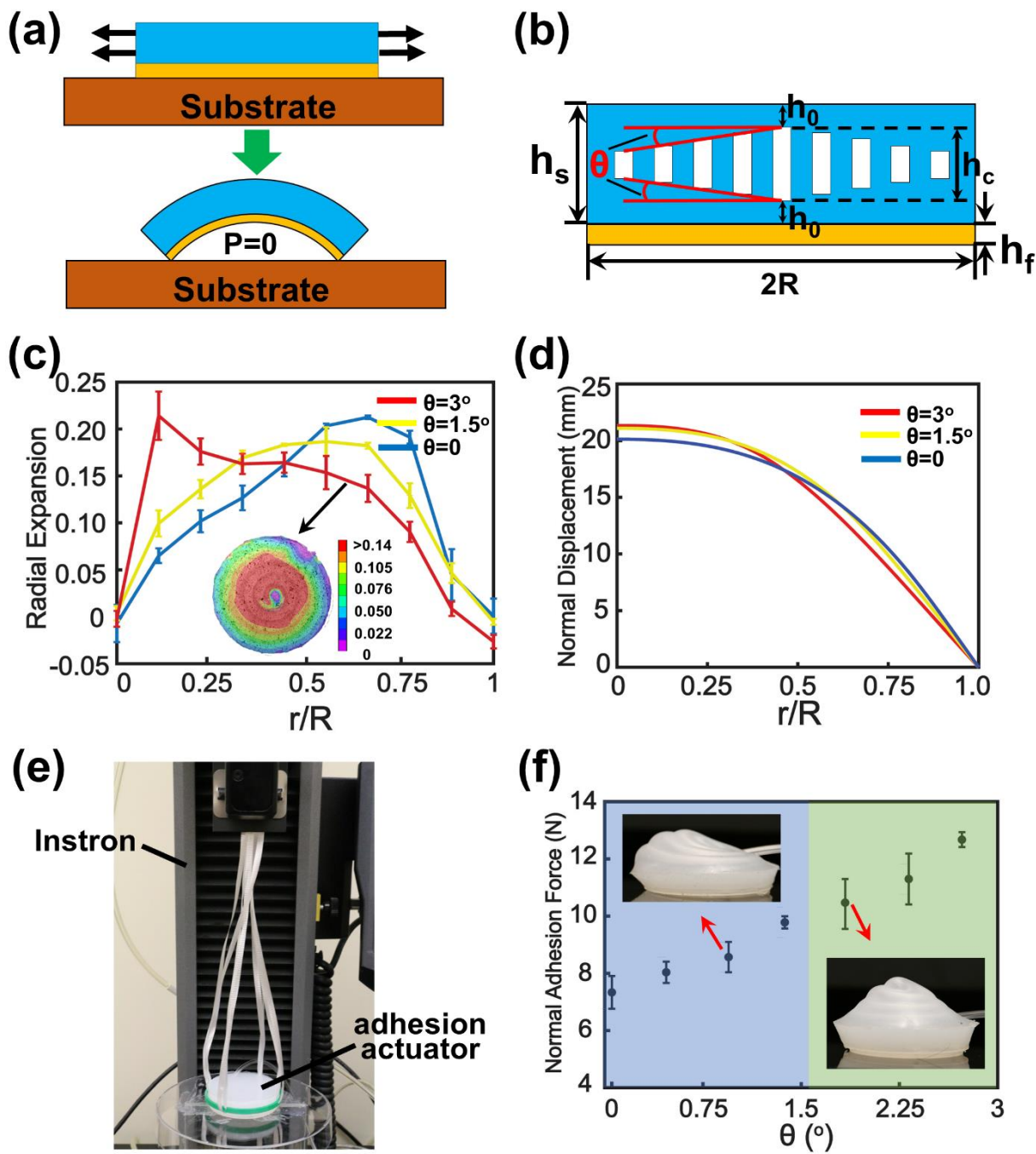

**Figure 4.** Design of bilayer doming-based soft adhesion actuator. (a) Schematic of the mechanism for switchable adhesion in the bilayer doming system upon expanding the top layer. top: adhesion-off state, bottom: adhesion-on state. (b) Schematic of the proposed bilayer doming model with pneumatic spiral channels in positive height gradient $\theta$. (c) DIC measured radial expansion rate of the actuators with $\theta=0°$, $\theta=1.5°$, and $\theta=3°$ upon 4mL inflation. The inset shows the radial strain contour in the actuator with $\theta=3°$. (d) Theoretically

predicted profiles of doming actuators with $\theta=0º$, $\theta=1.5º$, and $\theta=3º$. (e) Experimental set-up for measuring the pull-off force of the soft adhesion actuator. (f) The measured maximum normal adhesion force (pressurized at 40kPa) attached to acrylic surfaces as a function of $\theta$. Insets: collapsed asymmetric deformed shape (left) and axial symmetric deformed configuration (right).

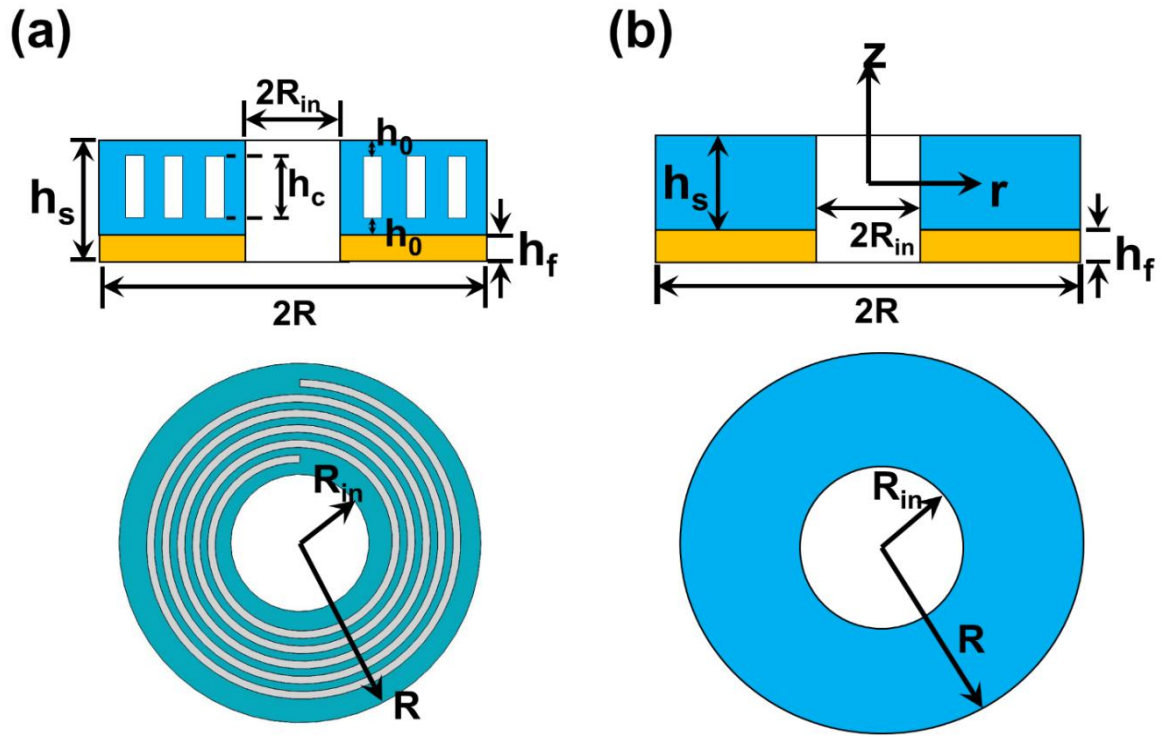

**Figure 5.** Schematic on the geometry of the proposed annulus bilayer doming actuator with pneumatic spiral channel (a) and the corresponding homogenized annulus bilayer doming model (b). The top figure shows the side-view and the bottom shows the top-view of the system

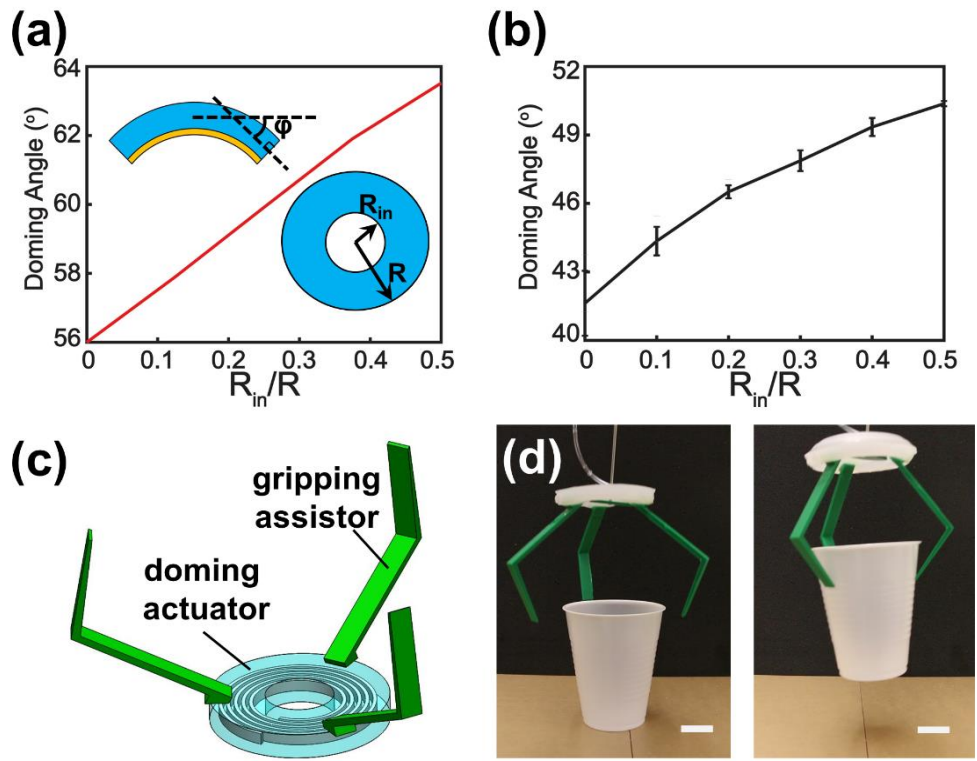

**Figure 6.** Design of bilayer doming-based soft gripper. (a) Theoretical doming angle vs. the normalized radius of the circular cut-out, $R_{in}/R$. (b) Measured doming angle as a function of $R_{in}/R$. (c) Schematic design of a gripper composed of a doming actuator (with spiral pneumatic channel) and polymeric grasping assistors (green) around the edge. (d) Demonstration of the proposed gripper grasping object. The scale bar is 20mm.

Supporting Information

**Design of Multifunctional Soft Doming Actuator for Soft Machines**
*Yichao Tang and Jie Yin[*]*

Y. Tang, Prof. J. Yin.
Applied Mechanics of Materials Laboratory
Department of Mechanical Engineering
Temple University
1947 North 12[th] Street, Philadelphia, PA 19122, USA
E-mail: jieyin@temple.edu

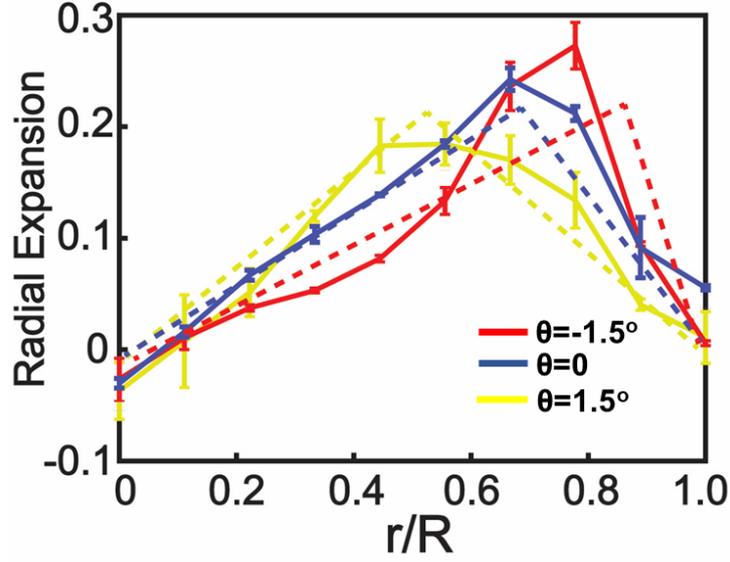

**Figure S1.** Fitting of the DIC measured radial expansion rate of the top layer with channel height gradient of $\theta$=-1.5º, $\theta$=0º and $\theta$=1.5º upon 4mL inflation. The solid lines denote the DIC measurements (Figure 3d) and the dashed lines represent the fitted curves.

The equation of fitted curve for $\theta$=1.5 º is:

$$\varepsilon_m(r) = \begin{cases} 0.0105r, & r \leq 0.5R \\ -0.0105r + 0.4, & r > 0.5R \end{cases}$$

The equation of fitted curve for $\theta$=0º is:

$$\varepsilon_m(r) = \begin{cases} 0.00752r, & r \leq 0.7R \\ -0.0175r + 0.667, & r > 0.7R \end{cases}$$

The equation of fitted curve for $\theta$=-1.5 º is:

$$\varepsilon_m(r) = \begin{cases} 0.00619r, & r \leq 0.85R \\ -0.0351r + 1.333, & r > 0.85R \end{cases}$$

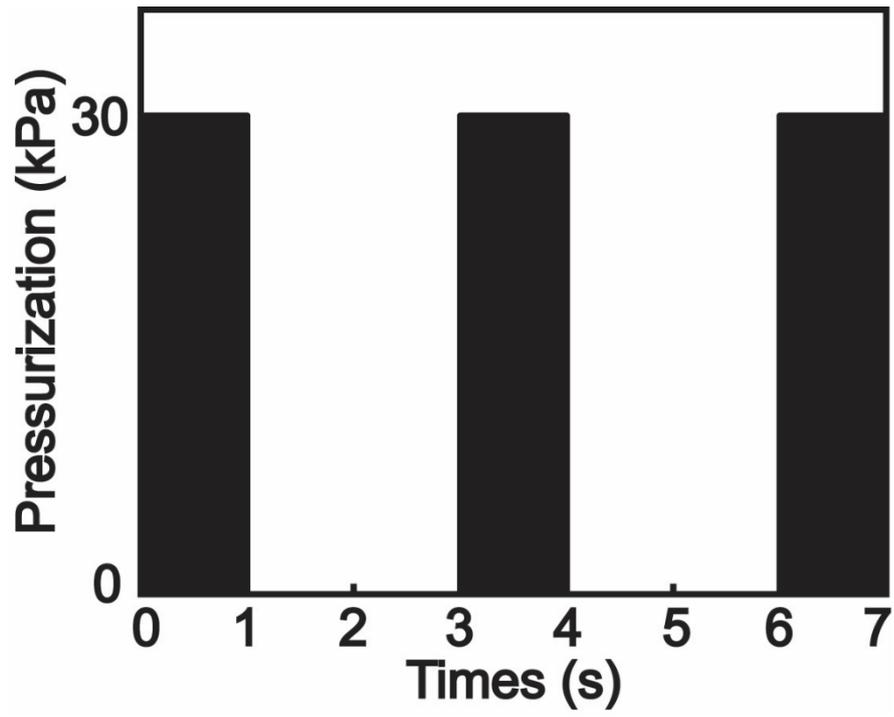

**Figure S2.** Actuation timing control pattern for the proposed swimming robot.

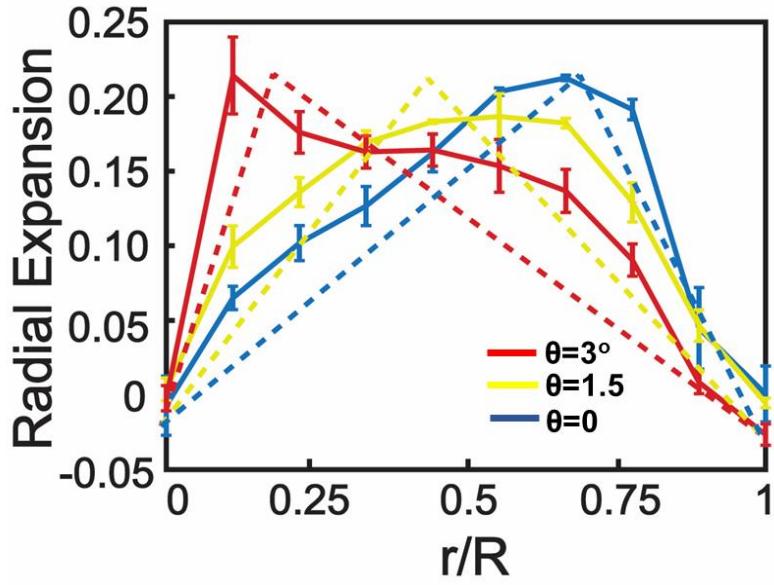

**Figure S3.** Fitting of the DIC measured radial expansion rate of the top layer with channel height gradient of $\theta=3°$, $\theta=1.5°$ and $\theta=0°$ upon 4mL inflation. The solid lines denote the DIC measurements (Figure 4c) and the dashed lines represent the fitted curves.

The equation of fitted curve for $\theta=3°$ is:

$$\varepsilon_m(r) = \begin{cases} 0.0263r, & r \leq 0.2R \\ -0.00658r + 0.25, & r > 0.2R \end{cases}$$

The equation of fitted curve for $\theta=1.5°$ is:

$$\varepsilon_m(r) = \begin{cases} 0.0105r, & r \leq 0.5R \\ -0.0105r + 0.4, & r > 0.5R \end{cases}$$

The equation of fitted curve for $\theta=0°$ is:

$$\varepsilon_m(r) = \begin{cases} 0.00752r, & r \leq 0.7R \\ -0.0175r + 0.667, & r > 0.7R \end{cases}$$

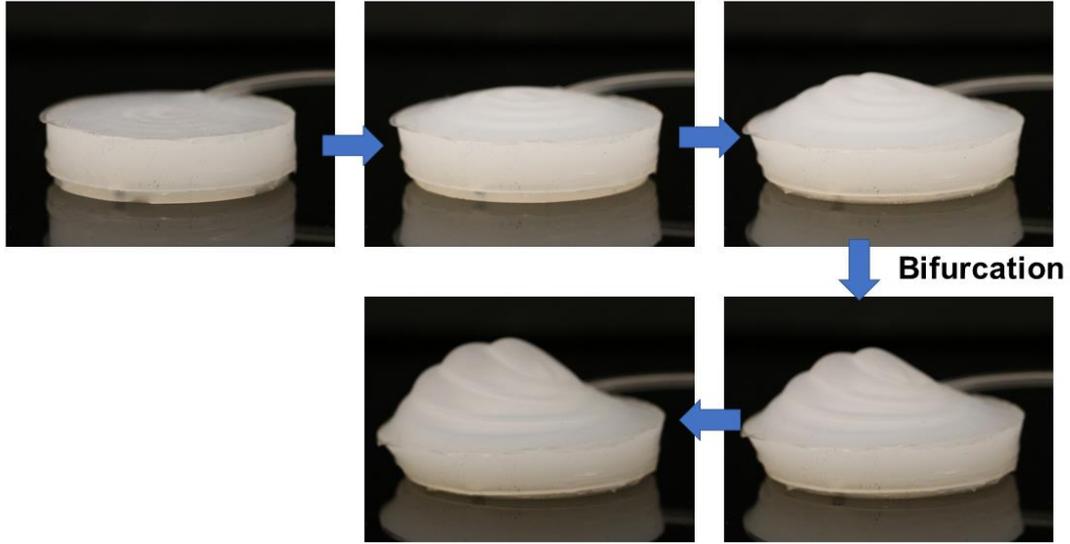

**Figure S4.** Deformation bifurcation of adhesion actuator with small channel height gradient of $\theta=0.9^{o}$ when attached to acrylics surface as the pressurization increases

**Modeling of Annulus Bilayer Doming**

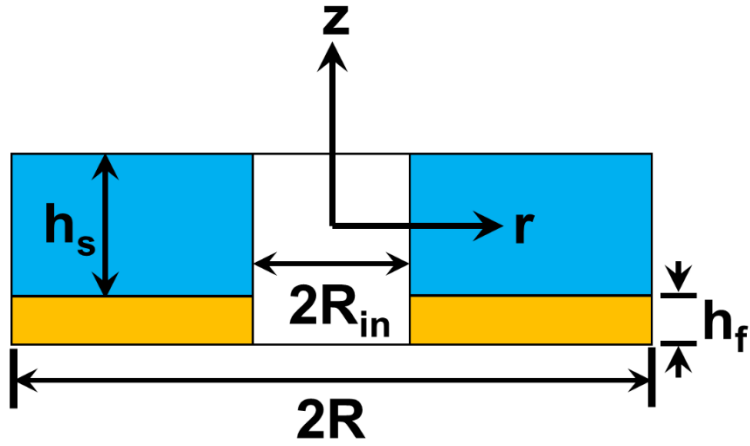

**Figure S5.** The schematic side view of annulus bilayer system.

The displacement of the deformed dome structure along the normal direction $z$-axis, $u_z$, can be obtained as

$$\frac{du_z}{dr} = 6\frac{E_f h_f}{1-v_f^2}\frac{1-v_s^2}{E_s h_s^2}\left[(1+vs)\frac{1}{r}\int_{R_{in}}^{r}\eta\varepsilon_m(\eta)d\eta\right] + Ar + B \quad (S1)$$

The displacement of the substrate (the blue layer in Figure S5) along the radial direction $u_r^s$ is:

$$u_r^s = \frac{E_f h_f}{1-\nu_f^2} \frac{1-\nu_s^2}{E_s h_s^2} \left[ (1+\nu_s)\frac{1}{r}\int_{R_{in}}^{r} \eta \varepsilon_m(\eta) d\eta \right] + Cr + D \quad (S2)$$

The displacement of the film (the orange layer in Figure S4) along the radial direction $u_r^f$ is:

$$u_r^f = u_r^s - \frac{h_s}{2}\frac{du_z}{dr} \quad (S3)$$

The axial forces in the substrate and film (denoted by blue and orange separately in Figure S5) are:

$$N_r = \frac{Eh}{1-\nu^2}\left[\frac{du_r}{dr} + \nu\frac{u_r}{r} - (1+\nu)\varepsilon_m\right] \quad (S4)$$

The bending moment of the substrate is:

$$M_r = \frac{E_s h_s^3}{12(1-\nu_s^2)}\left[\frac{d^2 u_z}{dr^2} + \frac{\nu_s}{r}\frac{du_z}{dr}\right] \quad (S5)$$

Constants A, B, C and D in Equation (S1) and Equation (S2) can be solved by applying the boundary conditions: the net force and net moment of the structure at the free edge of the bilayer (at $r=R$ and $r=R_{in}$) are zero:

$$\begin{cases} N_r^s + N_r^f = 0 \ @ \ r = R_{in} \\ N_r^s + N_r^f = 0 \ @ \ r = R \\ M_r - \frac{h_s}{2}N_r^f = 0 \ @ \ r = R_{in} \\ M_r - \frac{h_s}{2}N_r^f = 0 \ @ \ r = R \end{cases} \quad (S6)$$

where the superscripts "$s$" and "$f$" denote the substrate and film. By combining Equation (S1)-Equation (S6) we can solve constants A, B, C and D.

**Video S1**: Comparison of the real-time locomotion of the bilayer doming actuator based underwater soft robots with different channel height gradient of $\theta$=-1.5°, $\theta$=0° and $\theta$=1.5° by following the actuation control in Figure S2

**Video S2**: Proof-of-concept demonstration of the pick-up and release of a plastic cup through annulus bilayer doming-based soft gripper